\documentstyle[12pt,hpatex]{article}  
\input epsf
\begin{document}
\title{Textures and Cosmic Microwave Background \\[2mm] 
non--Gaussian Signatures}
\authors{Alejandro Gangui}
\address{ICTP -- International Center for Theoretical Physics,\\
P.~O.~Box 586, 34100 Trieste, Italy.}
\abstract{ 
We report briefly on a recent analysis of the texture--induced
CMB radiation three--point correlation function of temperature
anisotropies as predicted by an analytical model.  
We specialize our analysis to both large--scales (e.g., for 
{\sl COBE}--DMR, where we compare our prediction with the actual
four--year data) and  intermediate--scales. 
We show how the latter case puts strong constraints on the model parameters.
}
\section{Introduction}
The anisotropies in the Cosmic Microwave Background are a powerful 
test for models of structure formation in the universe. The
possibility of a departure from Gaussianity of the statistics of the 
anisotropies would disfavor standard inflation. It follows that it is
interesting to calculate the predictions texture make regarding
non--Gaussian signatures. 
It was recently proposed~\cite{ma95} a simple analytical model for the
computation of the ${\cal C}_\ell$'s from textures. The model exploits
the fact that in this scenario the microwave sky will show evidence of
spots due to perturbations in the effective temperature of the photons
resulting from the non--linear dynamics of concentrations of
energy--gradients of the texture field. 
The model of course does not aim to replace the full range numerical
simulations but just to show overall features predicted by textures in
the CMB anisotropies.
In fact the model leaves free a couple of parameters that are fed in
from numerical simulations, like the number density
of spots, $\nu$, the scaling size, $d_s$, and
the brightness factor of the particular spot, $a_k$, telling
us about its temperature relative to the mean sky temperature.
Texture configurations giving rise to spots in the CMB are assumed to
arise with a constant probability per Hubble volume and Hubble time.
In an expanding universe one may compute the surface probability
density of spots
\begin{equation}
dP = N(y) dy d\Omega, \quad \mbox{with} \quad 
N(y) = - {8 \nu \ln (2) \over 3} \left(2^{y/3} - 1 \right)^2, 
\end{equation}
where $\Omega$ stands for a solid angle on the two--sphere and 
the time variable $y(t) \equiv \log_{2} (t_0 / t)$ measures how
many times the Hubble radius has doubled since proper time $t$ 
up to now (e.g., for a redshift $z_{\rm ls}\sim 1400$ 
at last scattering we have
$y_{\rm ls} \simeq \log_{2}[(1400)^{3/2}] \simeq 16$).
In the present context the anisotropies arise from 
the superposition of the
contribution coming from all the individual spots $S_k$ produced from
$y_{\rm ls}$ up to now, and so, 
$\Delta T / T  = \sum_{k} a_{k} S_k(\theta_k, y)$, where the random
variable $a_{k}$ stands for the brightness of the hot/cold $k$--th 
spot with characteristic values to be extracted 
from numerical simulations~\cite{bo94}. 
$S_k (\theta_k, y)$ is the
characteristic shape of the spots produced at time $y$, where
$\theta_k $ is the angle in the sky measured with respect to the
center of the spot.
A spot appearing at time $y$ has typically a size
$\theta^s(y) \simeq d_s  ~ \theta^{\rm hor}(y)$, with
$\theta^{\rm hor}(y)$ the angular size of the horizon at
$y$, and where it follows that 
$\theta^s(y) = \arcsin \left( {0.5 d_s   \over 2^{y/3} - 1} \right)$.
Textures are essentially causal seeds and therefore the spots induced
by their dynamics cannot exceed the size of the horizon
at the time of formation, hence $d_s \leq 1$.
Furthermore the scaling hypothesis implies that the profiles satisfy
$S_k(\theta_k , y) = S(\theta_k / \theta^s(y))$.
From all this it follows a useful expression for the multipole 
coefficients, $a_{\ell}^{m} =
\sum_k a_k S^\ell_k (y) {Y_{\ell}^{ m}}^* (\hat\gamma _k )$,
with $S^\ell_k (y)$ the Legendre transform of the spot profiles.
At this point the ${\cal C}_\ell$'s are easily calculated~\cite{ma95}.
As we are mainly concerned with the three--point function we go on and
compute the angular bispectrum predicted within this analytical model,
which we find to be ($\bigl\langle  a^3 \bigr\rangle$ is the mean
cubic value of the spot brightness) 
\begin{equation}
\bigl\langle a_{\ell_1 }^{m_1} a_{\ell_2}^{ m_2} a_{\ell_3}^{ m_3}
\bigr\rangle =
\bigl\langle a^3 \bigr\rangle
\int dy N(y)  S^{\ell_1}(y) S^{\ell_2}(y) S^{\ell_3}(y)
\int d\Omega_{\hat\gamma} Y_{\ell_1}^{m_1} (\hat\gamma)
Y_{\ell_2}^{m_2}(\hat\gamma) Y_{\ell_3}^{m_3} (\hat\gamma) .
\end{equation}
Having the expression for the bispectrum we may just plug it in the
formulae for the full mean three--point temperature correlation 
function~\cite{ga94}. To make contact with experiments however we 
restrict ourselves to the collapsed case where two out of the three
legs of the three--point function {\it collapse} and only one angle,  
say $\alpha$, survives (this is in fact one
of the cases analyzed for the four--year {\sl COBE}--DMR data~\cite{wr96}).
The collapsed three--point function thus calculated,
$\bigl\langle C_3 (\alpha )\bigr\rangle$,
corresponds to the mean value expected in an
ensemble of realizations. However, as we can observe just one particular
realization, we have to take into account the spread of the
distribution of the three--point function values when comparing
a model prediction with the observational results. 
This is the well--known cosmic variance problem. 
We can estimate the range of expected values about the mean by the
{\it rms} dispersion
$\sigma_{CV}^2(\alpha ) \equiv
\bigl\langle C_3^2 (\alpha ) \bigr\rangle - \bigl\langle C_3 (\alpha )
\bigr\rangle^2$.
We will estimate the range for the amplitude of the
three--point correlation function predicted by the model by
$\bigl\langle  C_3 (\alpha ) \bigr\rangle \pm \sigma_{CV}(\alpha )$.
It has been shown~\cite{bo94} that spots generated from 
random field configurations of concentrations of
energy gradients lead to peak anisotropies 20 to 40\% smaller than
those predicted by the spherically symmetric self--similar texture 
solution. These studies also suggest an asymmetry
between maxima $\langle a_{\rm max} \rangle$ and minima
$\langle a_{\rm min} \rangle$ of the peaks as being due to the fact
that, for unwinding events, the minima are
generated earlier in the evolution 
(photons climbing out of the collapsing texture) 
than the maxima (photons falling in the collapsing
texture), and thus the field correlations are stronger for the maxima,
which enhance the anisotropies.
\section{Comparison with observations}
Let us now compute the predictions on the CMB
non--Gaussian features derived from the present analytical texture
model.
One needs to have the distribution of the spot brightness
$\{a_k\}$ in order to compute the mean values $\langle a^n \rangle$.
It is enough for our present purposes to take for all hot spots
the same $a_h > 0$ and for all the cold spots the same $a_c < 0$.
Then the $\langle a^n \rangle$
needed can readily be obtained in terms of $\langle a^2\rangle$ and
$x\equiv \langle a \rangle /\langle |a| \rangle$.
We fix $\langle a^2\rangle$ from the amplitude of the anisotropies
according to four--year {\sl COBE}--DMR~\cite{wr96}.
The other
parameter, $x$, that measures the possible {\it asymmetry} between hot
and cold spots, we leave as a free parameter.
We first consider the {\sl COBE}--DMR window function and, in order to take
into account the partial sky coverage due to the cut in the maps at
Galactic
latitudes $\vert b \vert < 20^\circ$, we multiply $\sigma_{CV}$ by
a factor $\sim 1.5$ in the numerical results
(sample variance).
Let us now compare with the data: Subtracting the dipole and for all
reasonable values of the asymmetry parameter $x$,
the data falls well within the
$\bigl\langle  C_3 (\alpha ) \bigr\rangle \pm \sigma_{CV}(\alpha )$
band, and thus there is good agreement with the observations.
However, the band for Gaussian distributed fluctuations
(e.g., as predicted by inflation) also
encompasses the data well enough, and
it is in turn included inside the texture predicted band.
This makes it impossible
to draw conclusions favoring one of the models~\cite{gamo96}.
\begin{figure}[t]
\centering
\leavevmode\epsfysize=10.5cm \epsfxsize=13.5cm
\epsfbox{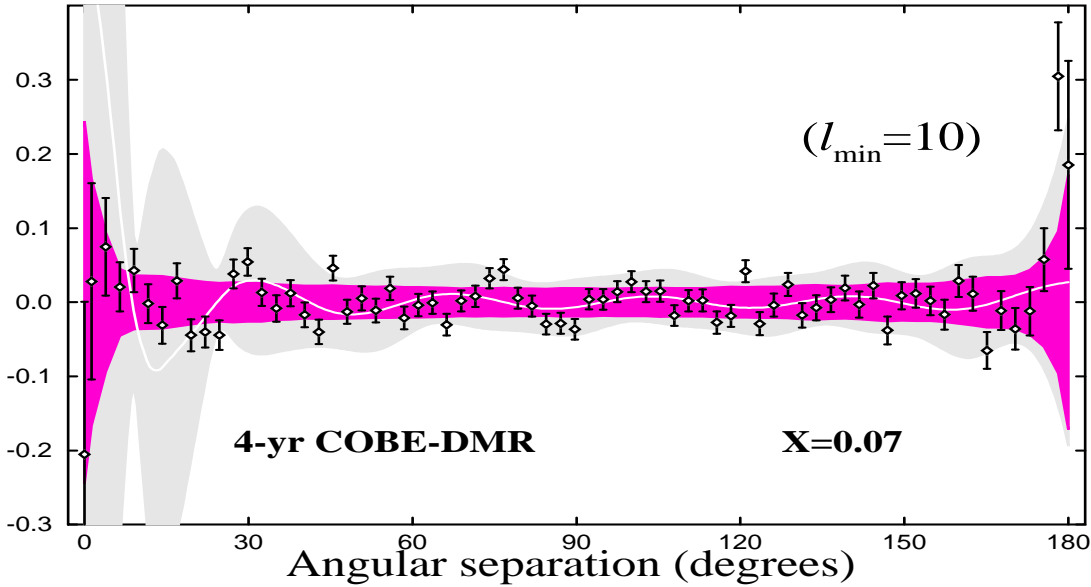}\\
\vspace{-1.7cm}
\parbox{16cm}{\caption[figure2]{\small
\baselineskip 12pt
{\sl 
The collapsed three--point function (white curve)
and the rms cosmic variance (grey band) which includes also the 
$\sim 50\%$ increment due to the sample variance, 
in units of $10^4 {\mu}{\rm K}^3$, 
as predicted by the analytical texture model. 
The asymmetry parameter is $x=0.07$, as suggested by simulations.
Also shown is the `pseudo--collapsed' three--point function
as computed from the analysis of the four--year {\sl COBE}--DMR data.
Error bars represent instrument noise while the dark band represents
the rms range of fluctuations due to a superposition of instrument
noise and cosmic and sample variance.
}
}}
\label{figuC-2}
\end{figure}
It is well known that the largest contribution to the cosmic variance
comes from the small values of $\ell$.
Thus, the situation may improve if one subtracts
the lower order multipoles contribution, as in a
$\ell_{min}=10$ analysis~\cite{hi95}.
In Figure 1 we show the analysis of the four--year {\sl COBE}--DMR
data evaluated from the 53 + 90 GHz combined map, containing power
from the $\ell = 10$ moment and up.
It is apparent that the fluctuations about zero correlation
(i.e., no signal) are too large for the instrument noise to be the
only responsible. These are however consistent with the range of
fluctuations expected from a Gaussian process (dark band).
What we want to see now is whether our analytical texture model
for the three--point function~\cite{gamo96} can do better when compared
with the data. In the same figure we show 
the collapsed three--point function
$\bigl\langle C_3 (\alpha ) \bigr\rangle$  (white curve) and the grey
band indicates the {\it rms} range of fluctuations expected from the
cosmic variance.
From this figure one may see qualitatively by eye that
(for some ranges of the angular separation better than for others,
of course) the data seems to follow `approximately'
the trend of the texture curves.

An experiment probing smaller angular scales than {\sl COBE} should
thus
be more appropriate to test non--Gaussian features in texture models.
As an example we compute the predictions for a three--beam subtraction
scheme experiment with window function at zero--lag
${\cal W}_\ell =
(1.5 - 2 P_\ell (\cos\theta) + 0.5 P_\ell (\cos 2\theta))^{1/2}
\exp(-\ell(\ell+1)\sigma^2/2)$,
where $\sigma = 0.64^\circ$ is the beam width and $\theta =
2.57^\circ$
is the chopping angle.
This window function is peaked at $\ell \sim 70$ and the range
of multipoles that significantly contribute to the three--point
function is from $\ell=10$ to $\ell=100$.
Hence we are still probing large enough scales and our results
are not strongly affected by the microphysics of the last scattering
surface.
We obtain for the skewness
$S\equiv \langle C_3(0)\rangle=(4.75 \pm 1.94) \times 10^4 \mu{\rm
K}^3$,
where the error band stands for the associated cosmic variance
$\sigma_{CV}^2(0)$,
for a value of 
$x=0.07$.
For comparison, the Gaussian adiabatic prediction is
$S_{\rm Gauss}= 0.06 \times 10^4 \mu{\rm K}^3$.
Thus, in this case even for reasonably small values of the asymmetry
parameter $x$ one such experiment can in principle distinguish between
inflation and texture predictions,
and thus put stronger constraints on the model parameters.

\noindent
{\bf Acknowledgments:}
I thank Silvia Mollerach for her collaboration in this work, and 
Gary Hinshaw for the courtesy of providing the 4-yr data, 
Ruth Durrer, David Wands 
for instructive conversations during these {\it journ\'ees},
and the organizers for their invitation to present our work.
I acknowledge partial funding from The British Council/Fundaci\'on
Antorchas, and thank Dennis Sciama for his continuous support.
 
\end{document}